\documentclass[graybox]{svmult}

\usepackage{mathptmx}       % selects Times Roman as basic font
\usepackage{helvet}         % selects Helvetica as sans-serif font
\usepackage{courier}        % selects Courier as typewriter font
\usepackage{type1cm}        % activate if the above 3 fonts are
\usepackage{makeidx}         % allows index generation
\usepackage{graphicx}        % standard LaTeX graphics tool
\usepackage{multicol}        % used for the two-column index
\usepackage[bottom]{footmisc}% places footnotes at page bottom
\usepackage{subfigure}
\usepackage{amsmath}
\usepackage{bm}
\usepackage{amsfonts}
\makeindex             % used for the subject index
\begin{document}

\title*{Modeling and Simulation of Inelastic Effects in Composite Cables}
\author{Davide Manfredo, Vanessa D\"orlich, Joachim Linn and Martin Arnold}
\institute{Davide Manfredo, Vanessa D\"orlich, Joachim Linn \at Fraunhofer ITWM, Fraunhofer Platz 1, 67663, Kaiserslautern, Germany \\ \email{[davide.manfredo,vanessa.doerlich,joachim.linn]@itwm.fraunhofer.de}
\and Martin Arnold \at Institute of Mathematics,
Martin Luther University Halle-Wittenberg,
Theodor-Lieser-Str. 5, 06120 Halle (Saale), Germany
\email{martin.arnold@mathematik.uni-halle.de}
}
\maketitle
%\abstract{Each chapter should be preceded by an abstract (10--15 lines long) that summarizes the content. The abstract will appear \textit{online} at \url{www.SpringerLink.com} and be available with unrestricted access. This allows unregistered users to read the abstract as a teaser for the complete chapter. As a general rule the abstracts will not appear in the printed version of your book unless it is the style of your particular book or that of the series to which your book belongs.
%Please use the 'starred' version of the new Springer \texttt{abstract} command for typesetting the text of the online abstracts (cf. source file of this chapter template \texttt{abstract}) and include them with the source files of your manuscript. Use the plain \texttt{abstract} command if the abstract is also to appear in the printed version of the book.}
%\texttt{
\abstract{The present work aims at describing hysteresis behaviour arising from cyclic bending experiments on cables by means of the Preisach operator. Pure bending experiments conducted in previous work show that slender structures such as electric cables behave inelastically and open hysteresis loops arise, with noticeable difference between the first load cycle and the following ones. The Preisach operator plays an important role in describing the input-output relation in hysteresis behaviours and it can be expressed as a superposition of relay operators. Here, we utilise data collected from pure bending experiments for a first approach. We introduce a mathematical formulation of the problem, and starting from the curvature of the cable specimen, we recursively define the Preisach plane for this specific case. Therefore, we derive a suitable kernel function in a way that the integration of such function over the Preisach plane results in the bending moment of the specimen.}
%}
\section{Introduction}
\label{sec:266_1}
\begin{figure}[h]
	\begin{center}
		\begin{subfigure}{}
			\includegraphics[width=0.11\textwidth]{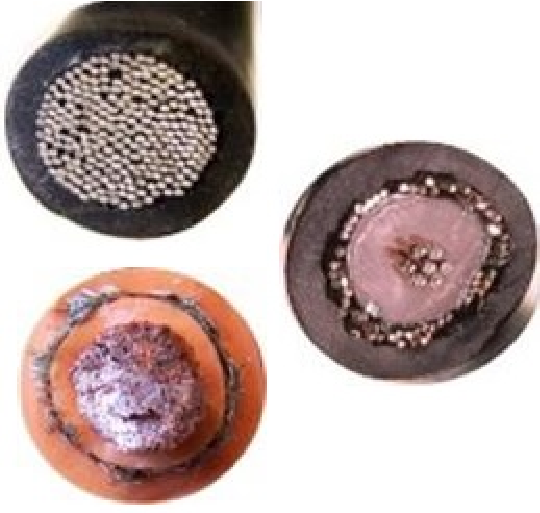}
		\end{subfigure}
		\begin{subfigure}{}
			\includegraphics[width=0.52\textwidth]{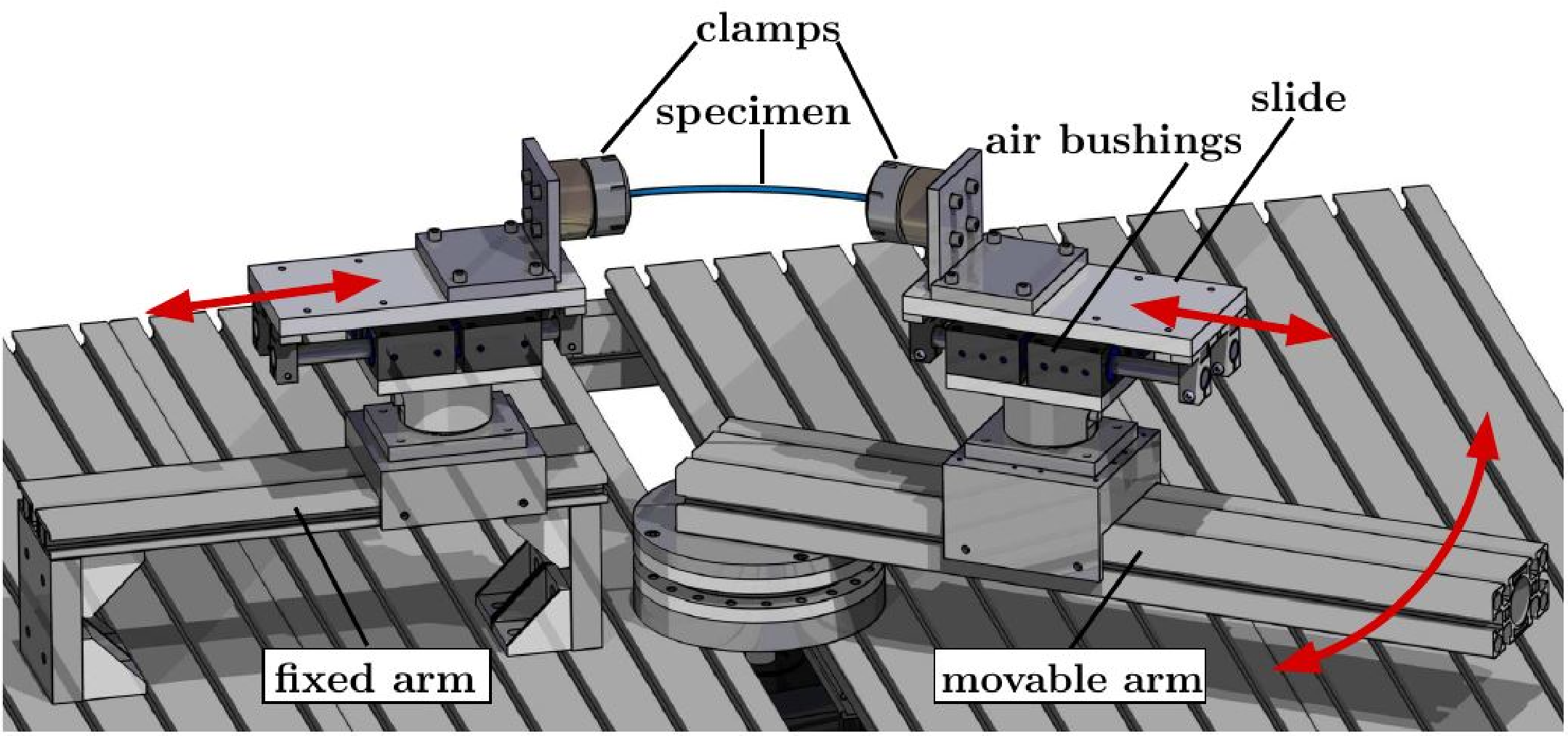}
		\end{subfigure}
		\begin{subfigure}{}
			\includegraphics[width=0.32\textwidth]{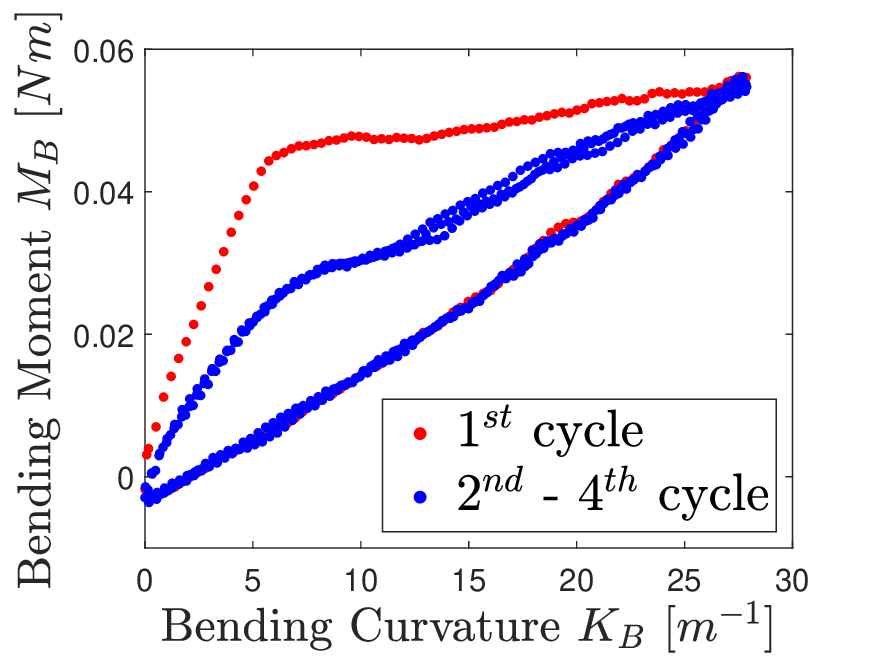}
		\end{subfigure}
		\caption{\textit{Left}: cross sections of different electric cables. \textit{Centre}: pure bending test rig. \textit{Right}: bending moment vs. bending curvature diagram measured in a pure bending experiment.}
		\label{fig:fig266_1}
	\end{center}
\end{figure}

Electric cables, as those shown in Fig. \ref{fig:fig266_1} \textit{left}, are complex objects due to their multi-material composition and their geometric properties. Consequently, different internal interaction effects occur and lead to an observed effective inelastic deformation behaviour of such cables. Cyclic bending experiments, Fig. \ref{fig:fig266_1} \textit{centre}, show open hysteresis loops with noticeable difference between the first load cycle and the following ones \cite{dorlich2018,dorlich2017}, as shown in Fig. \ref{fig:fig266_1} \textit{left}. In the framework of continuum mechanics, such deformation effects are modelled using suitable constitutive equations for specific material behaviour. In the presented work, we aim at modelling the observed behaviour on an abstract level using hysteresis operators. The choice of this mathematical framework is motivated by the ability of such operators to describe hysteresis phenomena with enough generality and without the need of a priori assumptions on the material behaviour.
\section{Hysteresis operators}
\label{sec:266_2}
As shown in \cite{brokate1996,visintin1994}, hysteresis operators are a well-studied topic with a variety of applications, mainly hysteresis effects arising from electric and magnetic phenomena. Such operators are normally used to describe the relation between two scalar time-dependent quantities that cannot be expressed in terms of a single-valued function.
\subsection{Relay operator}
Given any couple $(a_1,a_2)\in \mathbb{R}^2$ with $a_1<a_2$, we introduce the relay operator $\mathcal{R}_{a_1,a_2}$.
For any input function $v\in \mathcal{C}([0,T])$ and initial value $\xi\in\{\pm1\}$, the output\\ $w=\mathcal{R}_{a_1,a_2}[v]:[0,T]\rightarrow\{\pm1\}$ is equal to $-1$ if the input function value $v(t)$ crosses the threshold $a_1$ from above, and is equal to $+1$ if $v(t)$ crosses the threshold $a_2$ from below.\\
The relay operator can be interpreted as a switch operator between the values $-1$ and $+1$, with switching interval of width $a_2-a_1$ and centered in $(a_2 - a_1)/2$. A graphical representation of the relay operator is given in Fig. 2. A formal definition of the relay operator can be found in \cite{visintin1994}. 
\begin{figure}[h]
	\begin{center}
		\begin{subfigure}{}
			\includegraphics[width=0.317\textwidth]{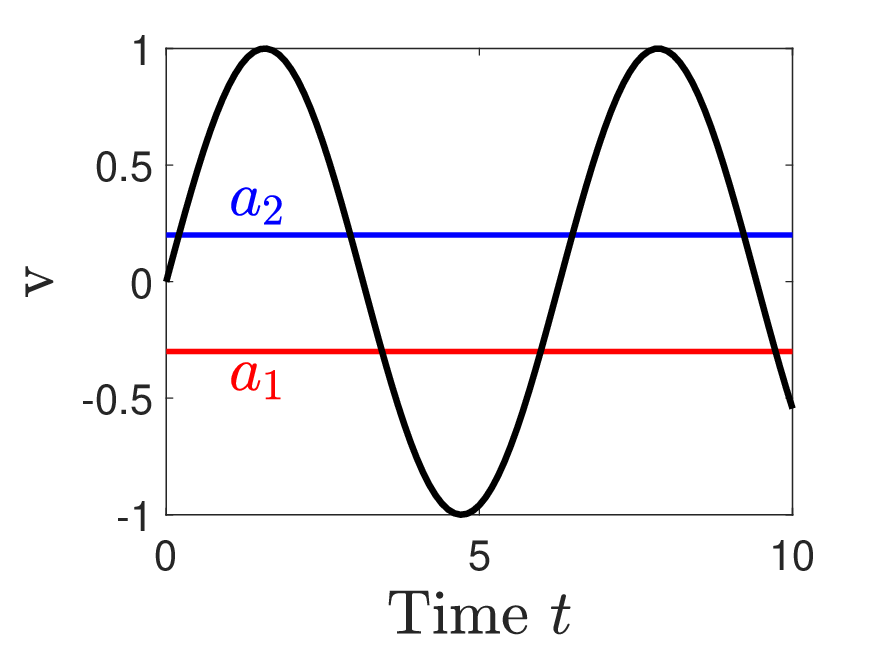}
			\label{fig:fig266_2a}
		\end{subfigure}
		\begin{subfigure}{}
			\includegraphics[width=0.317\textwidth]{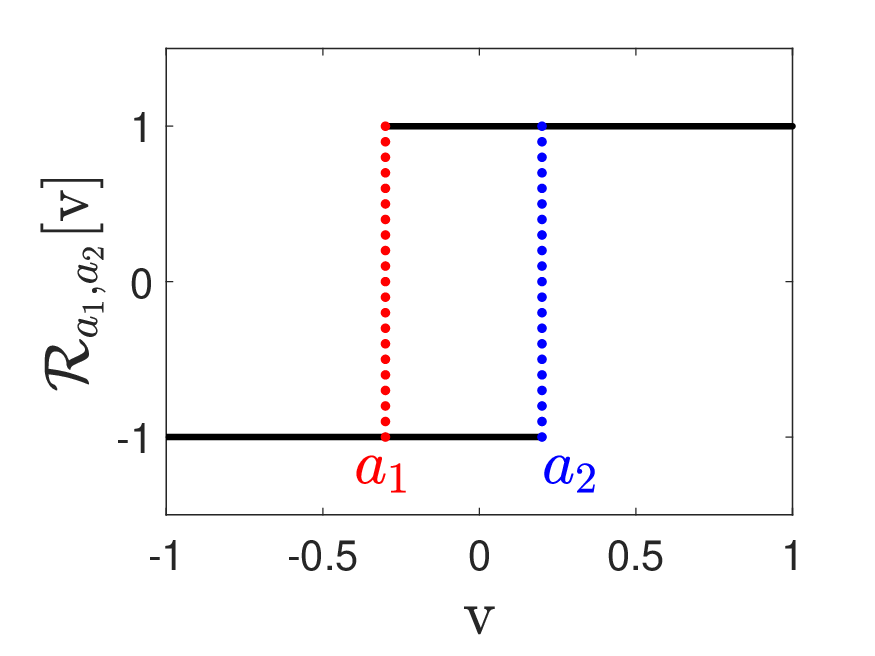}
			\label{fig:fig266_2b}
		\end{subfigure}
		\begin{subfigure}{}
			\includegraphics[width=0.317\textwidth]{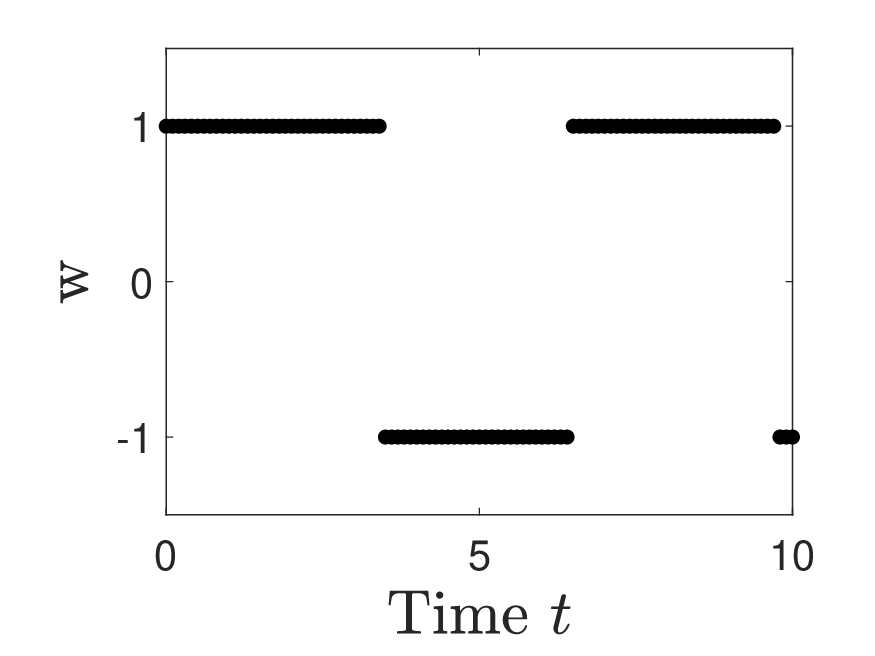}
			\label{fig:fig266_2c}
		\end{subfigure}
		\caption{\textit{Left}: input function $v(t)=\sin(t)$, with $t\in[0,10]$. \textit{Centre} diagram of the relay operator with $a_1=-0.3$ and $a_2=0.2$. \textit{Right}: output function $w(t)=\mathcal{R}_{a_1,a_2}[v](t)$, with initial value $\xi=+1$.}
		\label{fig:fig266_2}
	\end{center}
\end{figure}
\subsection{Preisach operator}
\begin{figure}[h]
	\begin{center}
	\begin{subfigure}{}
	\includegraphics[width=0.4\textwidth]{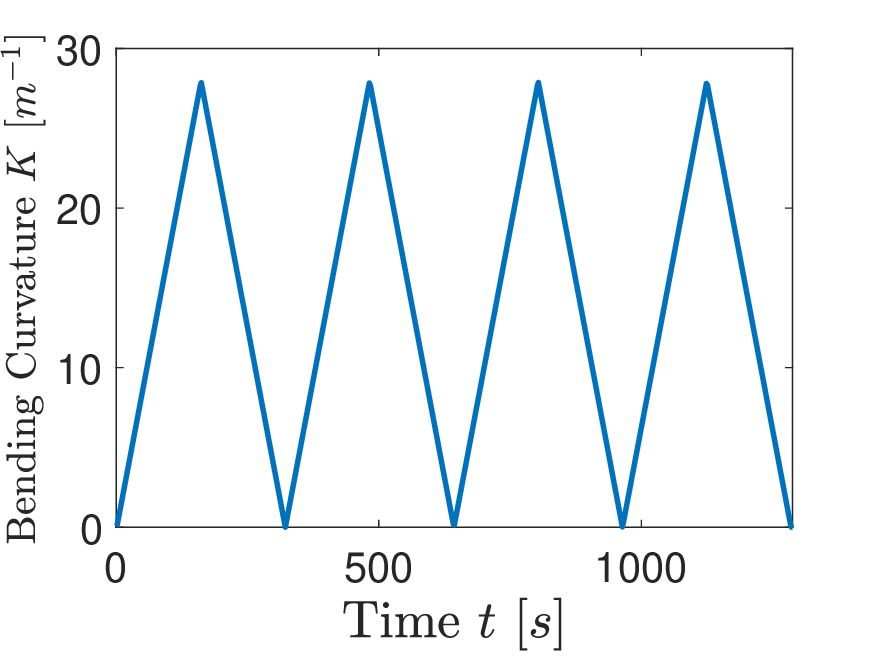}
	\end{subfigure}
	\begin{subfigure}{}
		\includegraphics[width=0.4\textwidth]{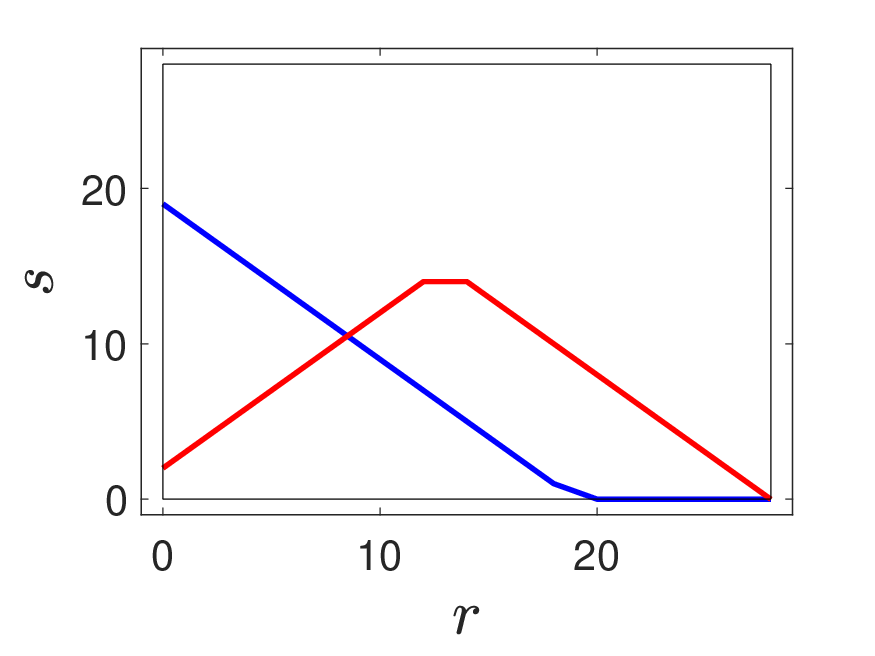}
	\end{subfigure}
	\begin{subfigure}{}
		\includegraphics[width=0.4\textwidth]{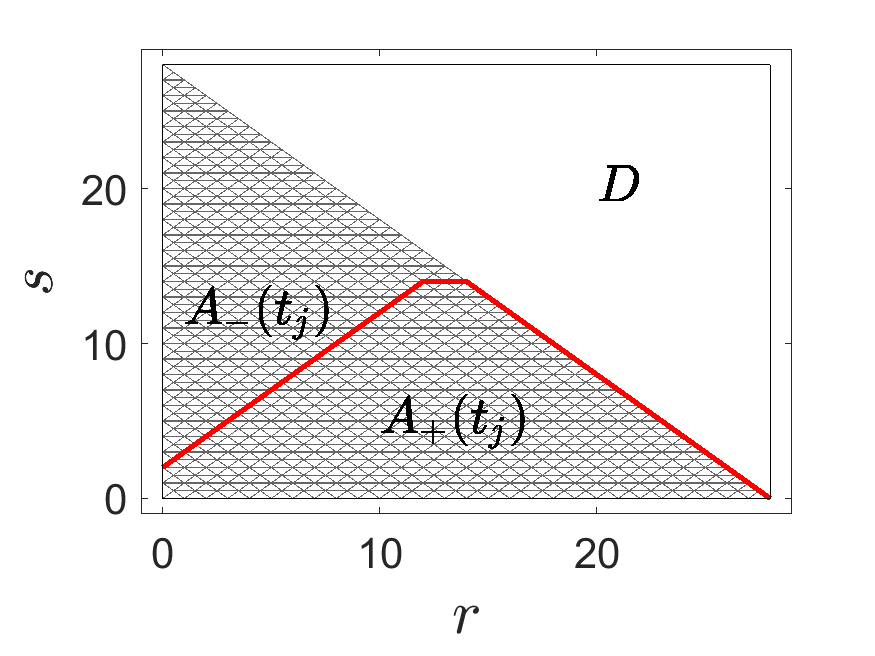}
	\end{subfigure}
	\caption{\textit{Top left}: input given as curvature vs. time. \textit{Top right} domain (black rectangle) included in the Preisach plane with two examples of memory curve. \textit{Bottom}: domain included in the Preisach plane with the triangulation and a memory curve for a given time $t_j$.}
	\label{fig:fig266_3}
	\end{center}
\end{figure}

The previously described relay operator is the "building block" of the Preisach operator. To be more precise, a superposition of relay operators multiplied by a suitable kernel function $\omega(r,s)$, assumed to vanish for large values of $|s|$ and $r$, defines the Preisach operator.  
\begin{equation}
w(t)=\mathcal{P}[v](t)=\int_0^{+\infty}\int_{-\infty}^{+\infty}\omega(r,s)\mathcal{R}_{s-r,s+r}[v](t)ds dr.
\label{eqn:eqn266_1}
\end{equation}
Here, $v$ and $w$ are respectively the input (Fig. \ref{fig:fig266_3} \textit{top left}) and the output function, $s$ and $r$ are the coordinates of the Preisach plane, and $\mathcal{R}_{s-r,s+r}$ is the relay operator.\\
If we consider an input function $v(t)$, for every time $t$ we determine the set
\begin{equation*}
A_{\pm}(t)=\{(r,s)\in\mathbb{R}_+\times\mathbb{R}:\mathcal{R}_{s-r,s+r}[v](t)=\pm 1 \}.
\end{equation*}
The union of such sets corresponds to the so-called Preisach plane, as will be explained in Section \ref{sec:266_3}. One can verify that the dividing line  $B(t)=\partial A_+(t)\cap\partial A_-(t)$, also called memory curve, at each time $t$ is the graph of a function which can be defined recursively and carries the total memory information present in the system at time $t$ \cite{brokate1996}. In Fig. \ref{fig:fig266_3} \textit{top right}, two examples of memory curves are shown. Using $\mathcal{R}_{s-r,s+r}[v](t)\in\{\pm 1\}$ and the definition of $A_\pm(t)$, \eqref{eqn:eqn266_1} can be rewritten as
\begin{equation*}
w(t)=\int_{A_+(t)}\omega(r,s)dsdr-\int_{A_-(t)}\omega(r,s)dsdr.
\end{equation*}
It should be noted that Preisach hysteresis operators provide a model for causal response \cite{visintin1994}, such that the output value $w(t)$ at time $t$ depends only on inputs $v(\bar{t})$ at past times $\bar{t}\leq t$. Thus, hysteresis loops can be computed by integrating a suitable kernel function $\omega(r,s)$ over a domain included in the Preisach plane.\\
\section{Problem formulation}
\label{sec:266_3}
As previously said, we aim at describing the input - output relation of bending curvature vs. bending moment by means of the Preisach operator, utilising data coming from a pure bending cyclic experiment. The available data are time $\{t_i\}_{1\leq i\leq T}$, bending curvature $\{K_{i}\}_{1\leq i\leq T}$ and bending moment $\{M_{i}\}_{1\leq i\leq T}$. Note that the values of time and bending curvature are prescribed by the experimental procedure, while the values of bending moment are measured.\\
Starting from the input function, for each time step $t_i$, we recursively define the Preisach plane, i.e. the sets $A_\pm(t_i)$ and the memory curve $B(t_i)$. Thus, our goal is to find $\omega(r,s)$ such that the following expression is minimised
\begin{equation}
\frac{1}{T}\sum_{i=1}^T\frac{1}{2}\left(M_{i}-\int\int_{A_+(t)}\omega(r,s)dsdr+\int\int_{A_-(t)}\omega(r,s)dsdr \right)^2.
\label{eqn:eqn266_2}
\end{equation}
To this end, we will take into account only a subset of the Preisach plane, namely the rectangle $[0,\max_{0\leq i\leq T}\{K_i\}]\times[0,\max_{0\leq i\leq T}\{K_i\}]$, since we assume $\omega(r,s)$ to vanish outside such domain.
Moreover, as shown in \cite{shirley2003}, we choose a tolerance $d$ to round the input values. Hence, we divide the part of the Preisach plane crossed by the memory curve $B(t)$ in $n-1$ triangles of equal area, such that at each time step, $B(t_i)$ lies on the edges of the triangles, see Fig. \ref{fig:fig266_3} \textit{bottom}. Now, we denote by $X\subset\mathbb{N}_+$ the set of indices given to the elements of the triangulation, by $e^m$, with $m\in X$, the triangles of the grid, and we define the sets
\begin{align*}
X_i=&\{m\in X| e^m \text{ below the memory curve at time }t_i\},\\
X\backslash X_i=&\{m\in X| e^m \text{ above the memory curve at time }t_i\}.
\end{align*}
As shown in Fig \ref{fig:fig266_3} \textit{top left}, we call $D$ the part of the Preisach plane that is never crossed by the memory curve.
We assume that the kernel function $\omega(r,s)$ is piecewise constant over each triangle of the mesh and over $D$, and we want to approximate the output as
\begin{equation*}
M_{i}\approx\sum_{m\in X_i}\int\int_{e^m}\omega(r,s)dsdr-\sum_{m\in X\backslash X_i}\int\int_{e^m}\omega(r,s)dsdr-c\quad i=1,...,T
\end{equation*}
$c$ being the constant value of the kernel function over $D$.
Now, we define the row vector $\boldsymbol{\Delta}_i=[\delta_i^1,...,\delta_i^{n-1},-1]$  for each time step $t_i$, where $\delta_i^m=1$ if $m\in X_i$ and $\delta_i^m=-1$ if $m\in X\backslash X_i$.
Calling $x^m=\int\int_{e^m}\omega(r,s)dsdr$, we have
\begin{equation}
\Delta=\begin{bmatrix}\boldsymbol{\Delta}_1\\ \vdots \\ \boldsymbol{\Delta}_T
\end{bmatrix}\in \mathbb{R}^{T\times n},\quad \boldsymbol{X}=\begin{bmatrix}x^1\\\vdots\\x^{n-1}\\c
\end{bmatrix}\in \mathbb{R}^n, \quad \boldsymbol{Y}\in\begin{bmatrix}
M_{1}\\\vdots\\M_{T}
\end{bmatrix}\in\mathbb{R}^T.
\label{eqn:eqn266_3}
\end{equation}
Hence, using \eqref{eqn:eqn266_2} and \eqref{eqn:eqn266_3}, the function to minimise is $f(\boldsymbol{X})=\frac{1}{2}\lVert\Delta\cdot\boldsymbol{X}-\boldsymbol{Y} \rVert^2$. In practice, one often deals with insufficient experimental data, yielding\\ $\text{rank}(\Delta)=q<\min\{T,n\}$ for the matrix $\Delta$. In order to compensate for the lack of data, we perform a singular value decomposition of the matrix $\Delta^T\Delta=USV^T$, where $S$ is a diagonal matrix, with $\text{rank}(S)=q$.\\
We extract $\hat{S}$, $\hat{U}$, $\hat{V}$ from $S$, $U$, $V$, respectively, by eliminating the rows and the columns of $S$ that are zero, and the corresponding columns of $U$ and $V$. Setting $\boldsymbol{X}=\hat{V}\boldsymbol{Z}$, the expression to minimise becomes $g(\boldsymbol{Z})=\boldsymbol{Z}^T\hat{S}\boldsymbol{Z}-\boldsymbol{Y}^T\Delta\cdot\hat{V}\boldsymbol{Y}$. It is easily verified, that once a minimiser $\boldsymbol{Z}^*$ of $g$ is found, then $\boldsymbol{X}^*=\hat{V}\boldsymbol{Z}^*$ minimises $f$.
\section{First results and conclusion}
\label{sec:266_4}
\begin{figure}[h]
	\begin{center}
		\begin{subfigure}{}
			\includegraphics[width=0.455\textwidth]{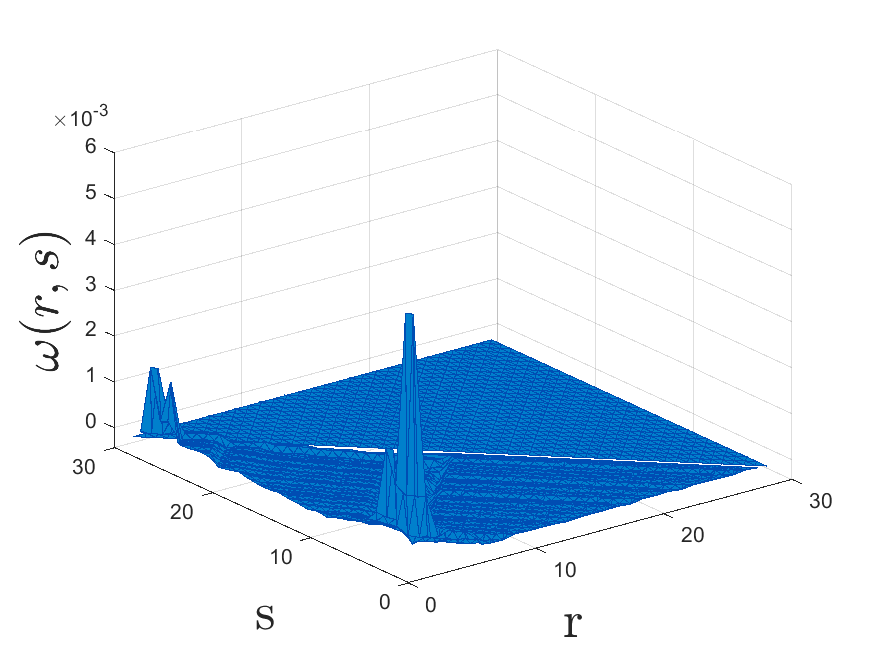}
			\label{fig:fig266_4a}
		\end{subfigure}
		\hspace{0.6cm}
		\begin{subfigure}{}
			\includegraphics[width=0.455\textwidth]{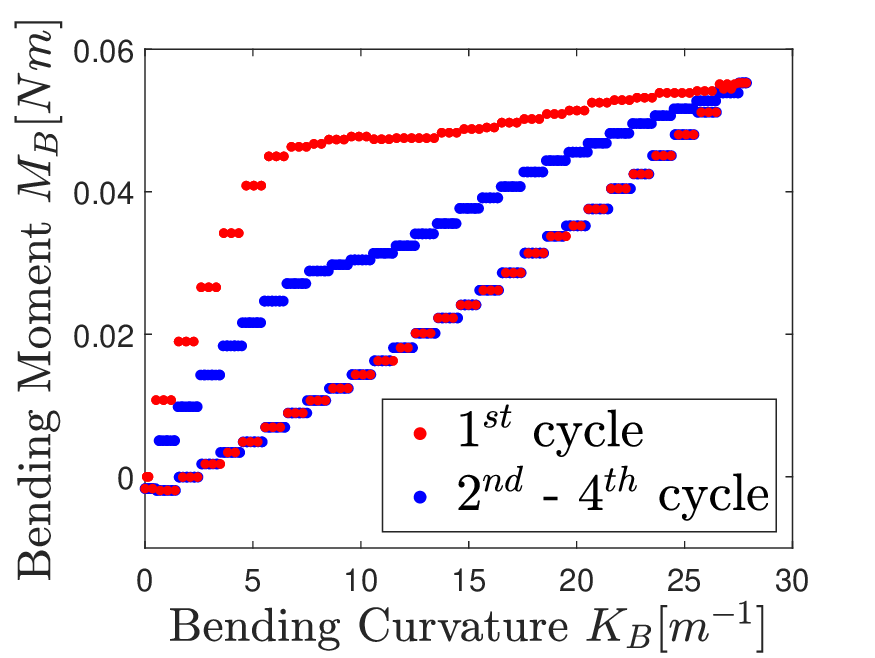}
			\label{fig:fig266_4b}
		\end{subfigure}
		\caption{\textit{Left}: kernel function obtained by the minimisation of $g$. \textit{Right}: estimated plot of bending moment vs. curvature obtained by means of the hysteresis operator.}
		\label{fig:fig266_4}
	\end{center}
\end{figure}

A minimiser $\boldsymbol{Z}^*$ of $g$ can be found using a Matlab routine such as "quadprog". In Fig. \ref{fig:fig266_4} \textit{left}, an approximation of the kernel function $\omega(r,s)$ is shown, and the integral of such kernel function over the domain included in the Preisach plane results in the diagram shown in Fig. \ref{fig:fig266_4} \textit{right}. Comparing the experimental data in Fig \ref{fig:fig266_1} \textit{right} with the diagram in Fig \ref{fig:fig266_4} \textit{right}, one can see that this approach describes the input - output relation as bending curvature vs. bending moment observed during the experiments quite well. One should note that the step-like behaviour of the diagram in Fig. \ref{fig:fig266_4} \textit{right} is due to the tolerance value $d$. However, the kernel function shows a highly nonlinear behaviour, and further work is necessary to investigate if its shape and properties are related to the physics of the studied phenomenon.\\
The Preisach operator is a very powerful and versatile tool to describe inelastic deformation behaviours of electric cables and the consequent open hysteresis loops arising from bending experiments. Moreover, such a mathematical tool captures the difference between load cycles very well and is relatively easy to implement. A more detailed study of the properties of the kernel function is necessary, with particular focus on its relation with the experimental data and the physics of the phenomenon.
\begin{acknowledgement}
This project has received funding from the European Union’s Horizon 2020 research and innovation programme under the Marie Skłodowska-Curie grant agreement No 860124.
\end{acknowledgement}

\end{document}